\documentclass{emulateapj}

\usepackage{epsfig}
\usepackage{graphicx}
\usepackage{float}
\usepackage[T1]{fontenc}
\usepackage[utf8]{inputenc}
\usepackage{lmodern}
\usepackage{amsmath}
\usepackage{placeins}
\usepackage{threeparttable}
\usepackage{booktabs}

\newcommand{\taub}{$\tau_{\text{b}}$}
\newcommand{\halphasurf}{$\Sigma_{\text{H}\alpha}$}
\newcommand{\lt}{$<$ }

\begin{document}
\title{Spatial Correlation Between Dust and H$\alpha$ Emission in Dwarf Irregular Galaxies\footnotemark[$\bigstar$]}
\author{Jimmy\footnotemark[1], Kim-Vy Tran\footnotemark[1], Am\'elie Saintonge\footnotemark[2], Gioacchino Accurso\footnotemark[2], Sarah Brough\footnotemark[3], Paola Oliva-Altamirano\footnotemark[3]\textsuperscript{,}\footnotemark[4], Brett Salmon\footnotemark[1], Ben Forrest\footnotemark[1]}
\footnotetext[$\bigstar$]{Based on VLT service mode observations (Programs 081.B-0649 and 083.B-0662) gathered at the European Southern Observatory, Chile.}
\footnotetext[1]{ George P. and Cynthia W. Mitchell Institute for Fundamental Physics and Astronomy, Department of Physics and Astronomy, Texas A\&M University, College Station, TX 77843, USA }
\footnotetext[2]{Department of Physics \& Astronomy, University College London, Gower Place, London WC1E 6BT, UK }
\footnotetext[3]{Australian Astronomical Observatory, PO Box 915, North Ryde, NSW 1670, Australia}
\footnotetext[4]{Centre for Astrophysics \& Supercomputing, Swinburne University of Technology, Hawthorn, VIC 3122, Australia}

\begin{abstract}

Using a sample of dwarf irregular galaxies selected from the ALFALFA blind HI-survey and observed using the VIMOS IFU, we investigate the relationship between H$\alpha$ emission and Balmer optical depth (\taub ).  We find a positive correlation between H$\alpha$ luminosity surface density and Balmer optical depth in { 8 of 11 at $\geq$ 0.8$\sigma$ significance (6 of 11 at $\geq$ 1.0$\sigma$)} galaxies.  Our spaxels have physical scales ranging from 30 to 80 pc, demonstrating that the correlation between these two variables continues to hold down to spatial scales as low as 30 pc.  Using the Spearman's rank correlation coefficient to test for correlation between \halphasurf\ and \taub\ in all the galaxies combined, we find $\rho = 0.39$, indicating a positive correlation at 4$\sigma$ significance.  Our low stellar-mass galaxy results are in agreement with observations of emission line regions in larger spiral galaxies, indicating that this relationship is independent of the size of the galaxy hosting the emission line region.  The positive correlation between H$\alpha$ luminosity and Balmer optical depth within spaxels is consistent with the hypothesis that young star-forming regions are surrounded by dusty birth-clouds.  

\end{abstract}

\maketitle

\section{Introduction}

H$\alpha$ has many advantages as a Star Formation Rate (SFR) tracer.  It does not have the degeneracies that are inherent in Spectral Energy Distribution (SED) fitting techniques, and it is less affected by extinction than SFRs based on the ultraviolet stellar continuum.  Finally, H$\alpha$ traces the instantaneous star formation rate, and is relatively insensitive to star formation history on longer timescales.  However, dust attenuates the light emerging from star-forming regions \citep{Kennicutt:12}.  Proper estimations of galaxy SFRs from H$\alpha$ luminosity rely on our understanding of the quantity, grain-size, and distribution of their dust \citep{Flaherty:07, Chapman:09, Shirley:11}.  

One must also be careful to make a distinction between attenuation of the stellar continuum and attenuation of ionized gas emission.  Throughout this work, we will focus on ionized gas attenuation, as it is most relevant to the Balmer series emission lines that we will be observing.  \citet{Reddy:15} found that the divergence between the reddening of the gas emission lines and the stellar continuum increases with increasing SFR.  They hypothesize two spatially independent components in galaxies, where stellar attenuation comes from the diffuse Interstellar Medium (ISM), and ionized gas attenuation comes from a higher dust content around localized birth clouds, which increases with increasing SFR.

For galaxies at high redshift (z $>$ 2), high spatial-resolution observations are not currently feasible, and therefore assumptions about the distribution of dust, and the dust laws, must be made when estimating the star formation rate.  Some studies at intermediate redshift (z $\sim$ 1-2) have found good agreement between local dust laws and high redshift dust laws \citep{Scoville:15,Reddy:15}, although others have suggested geometric or dust grain differences may cause different shapes to the dust law \citep{Kriek:13,Salmon:15b,Zeimann:15}.

Dwarf galaxies are expected to be Local Universe analogs of the young, gas-rich, star-forming galaxies observed at high redshift.  We can use these local galaxies to better understand the high-redshift population as they are both likely to contain little dust and low metallicity.  Unfortunately, little is currently known about the dust properties in low-metallicity dwarf galaxies due to the inherent difficulties involved in studying such low-luminosity systems \citep{Walter:07,Kreckel:13}.  ISM dust observations in dwarf galaxies are necessary to build more physically realistic dust models \citep{Jones:15}.

The Small Magellanic Cloud (SMC) is similar in mass and star formation rate to the dwarf galaxy population we will be studying, and it is the nearest and well-studied dwarf galaxy.  The SMC is known to have a steeper dust law curve than observed in the Milky Way \citep{Gordon:98, Misselt:99} or starburst galaxies \citep{Calzetti:00}.  Other studies have shown that dwarf irregular galaxies follow a dust law curve more similar to the SMC in their steepness \citep{Walcher:11}.  A steeper reddening curve implies that more blue light is preferentially attenuated/extincted.

Starburst galaxies have been shown to exhibit a flatter reddening curve \citep{Calzetti:94,Calzetti:00}.  Whether the differences between starburst galaxies and the SMC come from clumpy distributions of dust around stars or simply differences in the dust particles themselves \citep{Gordon:99} is uncertain.  \citet{Calzetti:97} supports the hypothesis that younger stars are more strongly attenuated by their birth clouds, while older stars are attenuated by the thin screen of dust in the ISM.

A spatial correlation between SFR and enhanced reddening in ionized regions has been observed in samples of larger spiral and elliptical galaxies \citep{Kreckel:13,Roche:15}.  The dusty birth-clouds that surround star-forming regions have lifetimes comparable to that of the massive stars that dominate H$\alpha$ emission line flux.  It is not yet known at which spatial scales this relationship begins to break down.

By combining the spatial distribution of H$\alpha$ and H$\beta$, we can map the influence of dust within each galaxy.  Balmer series transitions signify ionized gas regions within star-forming galaxies.  These emission line regions are known as HII regions and form near the most massive O stars, and typically have diameters on the scale of 0.1 pc \citep{Wood:89}.  The ionizing photons emitted from HII regions interact with the surrounding gas and dust to produce the emission lines of interest for this study.  The ratio of emission line fluxes for the Balmer series transitions have a known intrinsic value, given reasonable assumptions about the temperature and density of the gas.  We can study deviations from this expected value to estimate the influence of dust on light at the wavelengths of H$\alpha$ and H$\beta$.  It is typically assumed that H$\alpha$/H$\beta$=2.86 is the intrinsic ratio \citep{Osterbrock:89}, and that this is the lowest physically possible ratio of these two lines.

We present here, for the first time, spatial mapping of the H$\alpha$ emission and the Balmer optical depth in a sample of 11 dwarf irregular galaxies selected from the ALFALFA survey.  ALFALFA is a blind HI survey of the local universe (within 250 Mpc; \citealt{Haynes:11}).  The IFU observations of these low stellar-mass objects were taken using the VIMOS IFU spectrograph on the Very Large Telescope (VLT).  

The sample of dwarf irregular galaxies was first used in \citet{Jimmy:15} to study the fundamental metallicity relation as a function of HI-gas mass and SFR.  They estimate Oxygen abundances of each galaxy using the emission line ratios between H$\alpha$ and [NII].  They found that the dwarf galaxies were all sub-solar metallicity, as would be predicted by the mass-metallicity relation \citep{Jimmy:15}.  The median metallicity of all 11 galaxies was 12+log(O/H) = 8.22, which is comparable to the metallicity in the SMC (12+log(O/H) = 7.98; \citealt{Pagel:78}).  { They} also used dust corrected H$\alpha$ emission to estimate the SFR rate for each galaxy, and found a median SFR of 0.005 M$_\odot$ yr$^{-1}$, which is an order of magnitude lower than the SMC (0.037 M$_\odot$ yr$^{-1}$; \citealt{Bolatto:11}).  The similarities of the dwarf galaxies to the SMC leads us to assume an SMC dust law would be most appropriate for these galaxies.

We test to see if regions as small as (30 pc)$^2$ exhibit a correlation between higher H$\alpha$ luminosity and enhanced reddening.  Throughout this paper, we assume a Hubble constant of H$_0$ = 70 km s$^{-1}$ Mpc$^{-1}$.

\vspace{10 mm}
\section{Observations}
\label{Observations}

\subsection{IFU Spectroscopy}

Spectroscopic data of 28 dwarf irregular galaxies were taken using the VIMOS \citep{LeFevre:03} IFU spectrograph on the Very Large Telescope (VLT) located at Paranal Observatory.  These galaxies were selected from the ALFALFA survey to be nearby (D \lt 20Mpc) and low HI mass (M$_{\text{HI}} < 10^{8.2}$ M$_\odot$).  IFU spectroscopy is necessary to spatially map the reddening and H$\alpha$ emission throughout each galaxy.

The low stellar-masses and surface-brightnesses of dwarf irregular galaxies makes them difficult to observe.  Of the 28 observed galaxies, 11 have Balmer series emission lines greater than our amplitude over noise (AoN) cut of 3 for more than 20 spaxels and therefore are included in this study.  Integration times for the remaining 17 galaxies were insufficient to reach our target depth.  We find a small HI-gas mass bias between the detected and undetected galaxies in our original sample.  The median HI-gas mass for the undetected sample is 10$^{7.3}$ M$_\odot$ whereas we find an median of 10$^{7.6}$ M$_\odot$ for the detected sample.  The data were obtained starting on April 11, 2008 and ending on May 19, 2010 under program IDs 081.B-0649(A) and 083.B-0662(A).  Data were obtained using the VIMOS Low Resolution (LR) Blue Grism which has a wavelength range of 4000-6700 \AA\ and a spectral resolution of 5.3 \AA\ pixel$^{-1}$ (R $\sim$ 1000).

Using the LR Blue grism provides the full 54$^{\prime \prime}$x54$^{\prime \prime}$ field of view possible with VIMOS, which allows us to obtain spectra across the entire stellar disk of each galaxy in a single pointing.  Each object was observed using a 3 dither pattern, with each dither being integrated for 20 minutes.  Average seeing across all observations is 1.05" FWHM.

The LR blue grism provides both a wider spectral range, and a wider field of view when compared to the High Resolution blue grism.  The LR grism wavelength range allows for simultaneous observations of the emission lines from H$\beta$ ($\lambda$ = 4861 \AA ) to [NII] ($\lambda$ = 6583 \AA ).  The major drawback to using the low-resolution spectra ($\sim$5.3 \AA\ pixel$^{-1}$) is that we are unable to measure gas kinematics from the emission lines and the instrumental dispersion causes the H$\alpha$ and [NII] $\lambda \lambda$ = 6549,6583 \AA\ emission lines to blend together.  

\subsection{Data Reduction}

The spectroscopic data obtained with the VIMOS IFU is reduced from its raw form using the Reflex environment for ESO pipelines \citep{Freudling:13}.  The standard VIMOS template is used within the Reflex environment to produce the master bias and calibration frames containing the fiber traces and wavelength solution.  Many of the raw data and calibration frames contain a bright artifact across the surface of the chip, identified to be an internal reflection within the instrument.  This contamination interferes with the wavelength calibration routine within Reflex because it is often misidentified as a skyline, causing the spectrum to be shifted incorrectly.  To compensate, we disable the skyline shift in the calibration steps.  The wavelength solution without using the skyline shift provided by the ESO pipeline proved to be accurate within 5.3 \AA .  We also use the flux standardization routine within the Reflex VIMOS pipeline.  We then apply the calibration frames to the science frames to produce the fully reduced Row-Stacked Spectra (RSS).

The final output of the Reflex pipeline is four quadrants per observation.  We input these individual RSS quadrants into routines written in IDL and Python\footnotemark[1].  Further details of the process these scripts follow can be found in \citet{Jimmy:13}.  After the normalization and sky subtraction steps are completed, the two dimensional RSS are converted into three dimensional data cubes with two spatial dimensions and the wavelength on the third axis.  

\footnotetext[1]{Available publicly: http://jimmy.pink/\#code}

Once the data cubes have been built, the individual dithers (typically 3 per galaxy) are stacked using a 5$\sigma$ clipped mean.  We use the AoN to select for spaxels containing sufficient emission line flux for Balmer optical depth measurements.  AoN is defined as the amplitude of the emission line divided by the noise after the linear offset is subtracted.  An AoN threshold of 3 is used to select spaxels for analysis.  Only spaxels which pass the AoN cut in both the H$\alpha\ \lambda = 6563$\AA\ and H$\beta$ $\lambda = 4861$\AA\ emission lines are included in our analysis.

\begin{figure}
\epsfig{ file=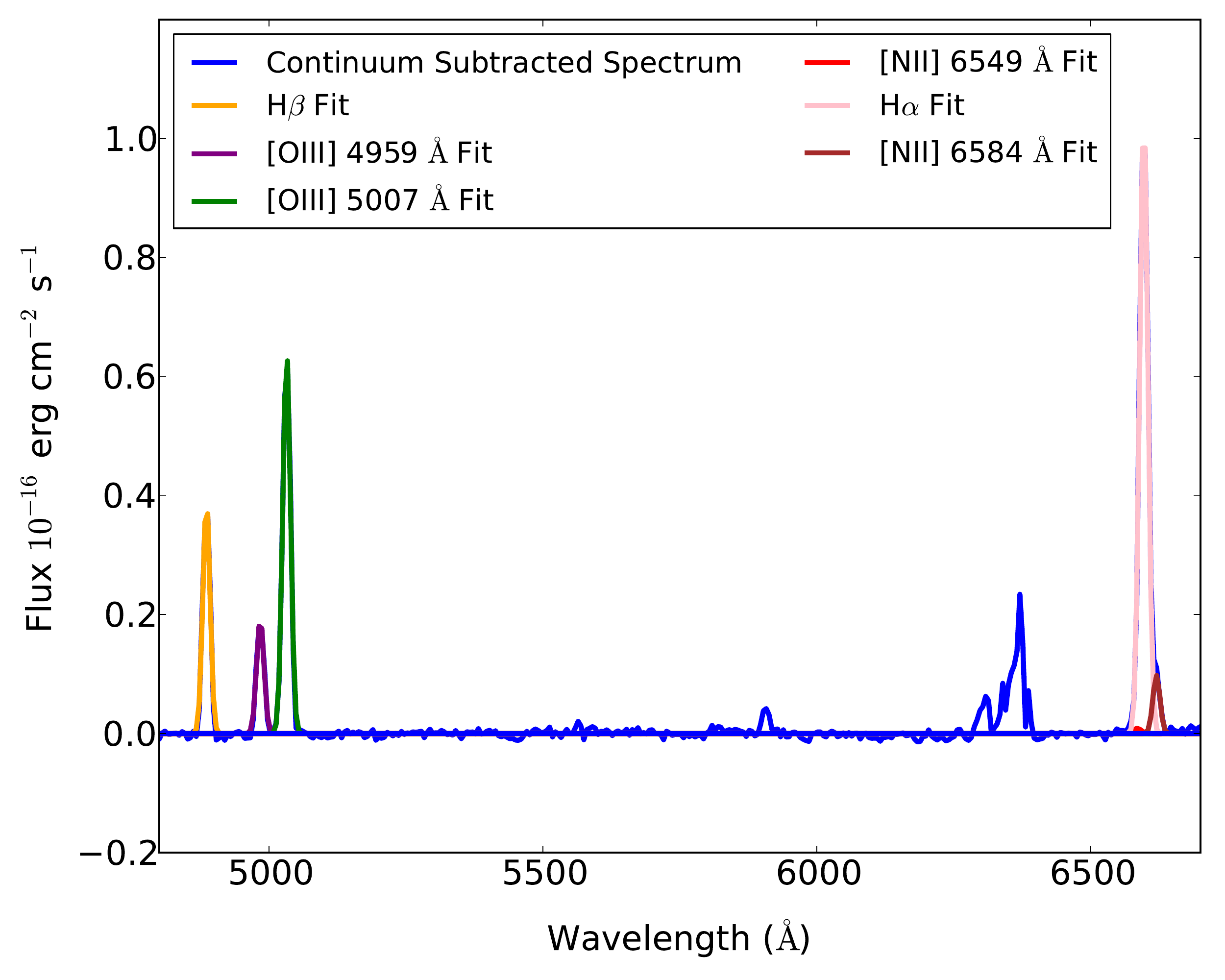, scale=0.35}
\caption[Example Result of Python-based Gaussian Fitting Routines]{Representative example spectra from a bright spaxel in AGC221000 showing the results of our Python-based Gaussian fitting routines.  The [NII] and H$\alpha$ emission lines are blended due to the instrumental resolution (5.3 \AA ).  Our Python-based Gaussian fitting routines are able to deconvolve the three emission lines into their constituent parts, as can be seen by the Gaussian curves plotted.  The flux in the red [NII] 6549 \AA\ emission line is negligible.  The anomalous feature near 6400 \AA\ is the result of internal reflections within the instrument.} 
\label{line_fit_output}
\end{figure}

All spaxels which pass our AoN cut are then fed into Python-based Gaussian fitting routines to measure the emission line fluxes.  To obtain the integrated spectrum, we sum the spectra from each spaxel that passes the AoN cut to produce a single spectrum.  The procedure that our Python-based Gaussian fitting routines follow is outlined in \citet{Jimmy:15}, but we will summarize it here.  A linear fit to the continuum is performed in the area immediately surrounding the triplets of emission lines on the blue and red ends of the spectrum (Figure \ref{line_fit_output}).  We fit 3 Gaussians simultaneously using the gaussfitter\footnotemark[2] routines, fitting the continuum and the Gaussians on the blue end and the red end independently.  As described in \citet{Jimmy:15}, the Python-based Gaussian fitting routines are able to successfully deblend the H$\alpha$ and [NII] emission lines.

We do not include a fit to the Balmer absorption in our procedures.  Our observations are insufficient to reliably estimate the stellar continuum for accurate absorption measurements.  However, it is likely that Balmer absorption effects are negligible and well within our uncertainties \citep{Rosa-Gonzalez:02,Reddy:15}.  Uncertainties on our flux measurements are estimated using a Monte Carlo technique with 1000 iterations.  

\footnotetext[2]{written by Adam Ginsburg http://www.adamgginsburg.com/pygaussfit.htm}

\subsection{Artificial Galaxy}

To ensure that any measured correlation between H$\alpha$ luminosity and enhanced reddening is robust, we have created artificial spectra to represent a galaxy with a constant Balmer optical depth and an exponentially decreasing line flux.  This will help us test for possible biases in our data.  We test whether or not noise is driving any observed relation between H$\alpha$ and the dust content.  Also since we are using the ratio between H$\alpha$ and H$\beta$ as a tracer of the dust, we must ensure that any measured correlation between H$\alpha$ and the dust content is not the result of H$\alpha$ flux also being used to measure the dust content.

The artificial galaxy is created using emission line ratios obtained for NGC 784 A in the \citet{Berg:12} low-luminosity galaxy sample.  This galaxy was chosen because the measured Balmer decrement (2.89; \citealt{Berg:12}) is closest to the range of Balmer decrements in this sample.  We scale the brightest central spaxel in the artificial galaxy to have an H$\beta$ emission line flux of $30.0 \times 10^{-16} \text{ erg cm}^{-2} \text{ s}^{-1} \text{ \AA} ^{-1}$, which corresponds to the highest H$\beta$ flux in the brightest spaxel of all galaxies in the dwarf IFU sample.  Each emission line is given a dispersion equal to the average instrumental dispersion of 7.5 \AA .  This galaxy was created to have an exponential decline in both H$\alpha$ flux and H$\beta$ flux with radius, but to keep the H$\alpha$/H$\beta$ ratio constant throughout.

To create the artificial galaxy, we first take a pipeline reduced data cube from one of the galaxies in our sample that was too faint to be detected by the IFU spectrograph (AGC220261) and find a region of the field of view which contains only sky with little instrumental artifacts.  We also check to ensure that none of the galaxy that was originally intended to be observed is located near this region.  We then generate emission lines to represent an artificial galaxy and place them within this region.  The artificial galaxy is then run through the full data analysis procedures described in \citet{Jimmy:15} to measure the recovered emission line ratios.

\vspace{10 mm}

\section{Results}

\begin{table*}
\centering\footnotesize
\caption{Integrated Properties of Dwarf Galaxies}
\label{paper2_table}\medskip
\begin{threeparttable}
\begin{tabular}{l r r r r r r r r}
\toprule
Galaxy & RA (J2000)& Dec (J2000)& Distance$^a$  & pc/spaxel  & HI Mass$^a$ & Stellar Mass$^b$ & Metallicity$^b$ & Dust Corrected SFR$^b$\\
AGC\# & hh:mm:ss.s & $\pm$hh:mm:ss & Mpc ($\pm$ 2.43) & ($\pm$ 8) & log(M$_\sun$) & log(M$_\sun$) & 12+log(O/H) & log(M$_\sun$yr$^{-1}$) \\
\midrule
191702 & 09:08:36.5 & +05:17:32 & 8.7 & 28 & 7.74 $\pm$ 0.18 & 6.67 $\pm$ 0.61 & 7.94 $\pm$ 0.13 & -2.76 $\pm$ 0.34\\
212838 & 11:34:53.4 & +11:01:10 & 10.3 & 33 & 7.60 $\pm$ 0.19 & 6.94 $\pm$ 0.56 & 8.22 $\pm$ 0.11 & -2.73 $\pm$ 0.32\\
220755 & 12:32:47.0 & +07:47:58 & 16.4 & 52 & 7.18 $\pm$ 1.20 & 7.76 $\pm$ 0.43 & 8.49 $\pm$ 0.16 & -2.51 $\pm$ 0.64\\
220837 & 12:36:34.9 & +08:03:17 & 16.4 & 52 & 7.41 $\pm$ 0.54 & 8.78 $\pm$ 0.46 & 8.56 $\pm$ 0.13 & -2.20 $\pm$ 1.28\\
220860 & 12:38:15.5 & +06:59:40 & 16.4 & 52 & 7.22 $\pm$ 1.39 & 7.57 $\pm$ 0.42 & 7.82 $\pm$ 0.13 & -2.14 $\pm$ 0.14\\
221000 & 12:46:04.4 & +08:28:34 & 16.5 & 53 & 7.46 $\pm$ 0.83 & 8.35 $\pm$ 0.44 & 8.35 $\pm$ 0.05 & -1.65 $\pm$ 0.08\\
225852 & 12:59:41.9 & +10:43:40 & 16.6 & 53 & 7.68 $\pm$ 0.53 & 7.57 $\pm$ 0.42 & 8.27 $\pm$ 0.22 &  -2.53 $\pm$ 0.38\\
227897 & 12:50:04.2 & +06:50:51 & 16.6 & 53 & 7.44 $\pm$ 0.89 & 6.58 $\pm$ 0.45 & 7.67 $\pm$ 0.42 & -2.88 $\pm$ 0.24\\
221004 & 12:46:15.3 & +10:12:20 & 16.7 & 53 & 7.66 $\pm$ 0.55 & 7.98 $\pm$ 0.30 & 8.35 $\pm$ 0.16 & -2.23 $\pm$ 0.22\\
202218 & 10:28:55.8 & +09:51:47 & 19.6 & 63 & 7.75 $\pm$ 0.50 & 8.12 $\pm$ 0.38 & 8.22 $\pm$ 0.11 & -1.72 $\pm$ 0.29\\
225882 & 12:03:26.3 & +13:27:34 & 23.6 & 76 & 8.15 $\pm$ 0.30 & 7.06 $\pm$ 0.35 & 7.95 $\pm$ 0.18 &  -2.33 $\pm$ 0.05\\
\bottomrule
\end{tabular}
\begin{tablenotes}[para,flushleft]
$^a$ Values obtained from the ALFALFA $\alpha$.40 catalog \citep{Haynes:11}.  Uncertainties in the distances are dominated by the local velocity
dispersion measured by \citealt{Masters:05}. $^b$Integrated measurements obtained from \citet{Jimmy:15}.   These SFR estimations have been corrected for reddening due to dust.
\end{tablenotes}
\end{threeparttable}
\end{table*}

Using VIMOS IFU spectroscopy, we are able to produce two-dimensional spatial maps of the physical properties of each galaxy.  Using the distances reported from the ALFALFA survey (Table \ref{paper2_table}), we convert the line flux into luminosity measurements to produce spatial mappings of the H$\alpha$ and H$\beta$ luminosities.  We also map the Balmer optical depth, defined as \begin{equation} \tau_{\text{b}} = \text{ln} \left\{ \frac{\text{H}\alpha/\text{H}\beta}{2.86} \right\} \label{taub}\end{equation} utilizing both the H$\alpha$ and H$\beta$ flux.  The H$\alpha$, H$\beta$, and \taub\ maps show all the same spaxels, due to the AoN selection criteria that both lines must be detected.  Inset within each frame of the spatial distributions is the integrated value, measured by stacking all of the spectra for each spaxel shown.  Each figure also shows the artificial spectra created to simulate a galaxy with a constant Balmer optical depth.  We have applied the same analysis to the artificial galaxy as the observed galaxies throughout.

\subsection{Balmer Line Luminosities}

\begin{figure*}
\epsfig{ file=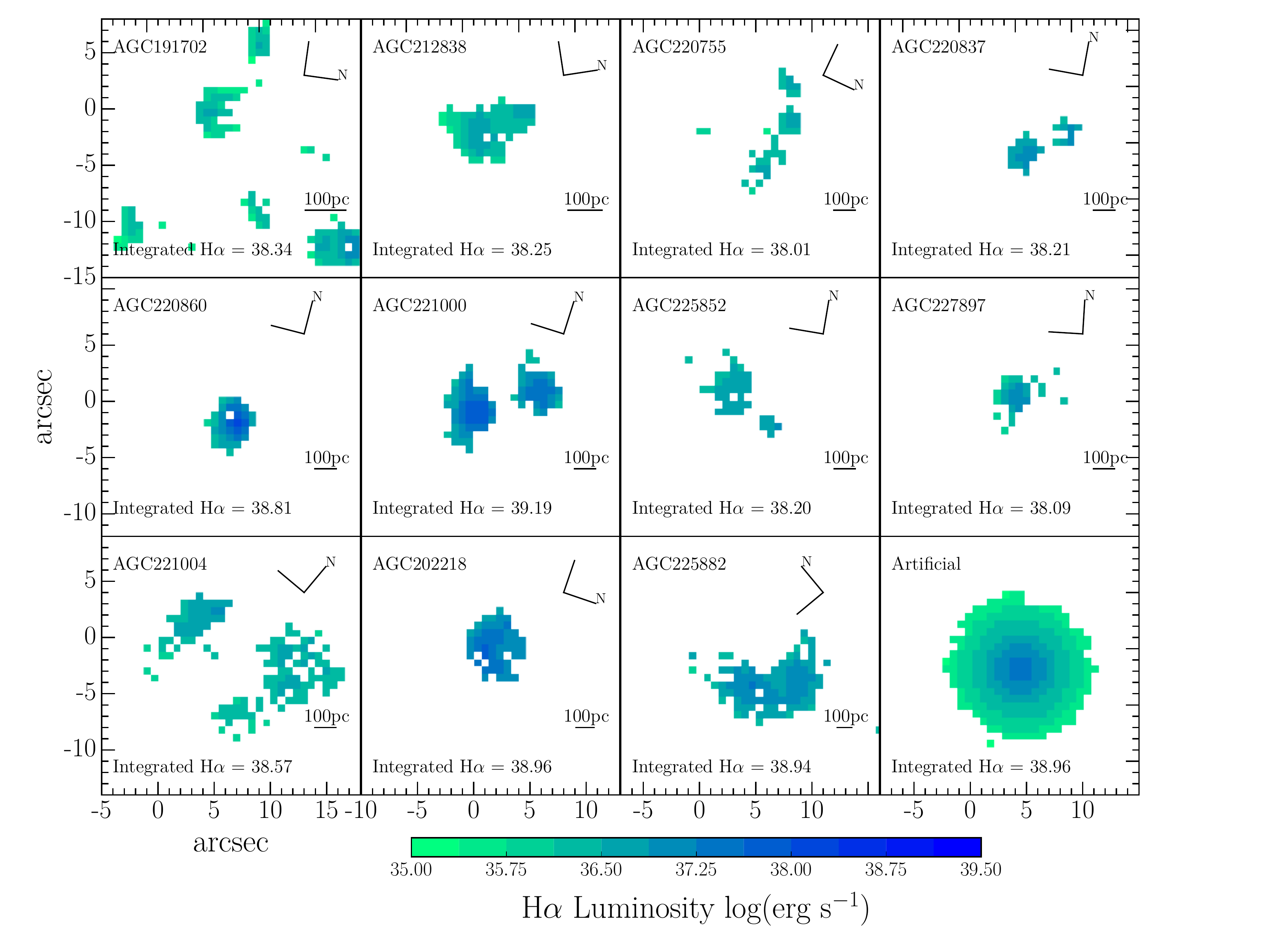, scale=0.6}
\caption[Spatial Map of H$\alpha$ Luminosity]{Spatial mapping of the H$\alpha$ flux for each galaxy.  Inset within each frame is the log H$\alpha$ flux (erg s$^{-1}$) measured when summing the spaxels together to form a single effective spectrum.  Peaks in the H$\alpha$ flux appear to be spatially correlated with peaks in the Balmer optical depth as seen in Figure \ref{balmer}.  Also shown in the lower right panel is the artificial galaxy created with an exponential decline in H$\alpha$ flux as a function of radius.}
\label{h_alpha}
\end{figure*}

\begin{table*}
\centering\footnotesize
\caption{Results from Integrated Spectra}
\label{results_table}\medskip
\begin{threeparttable}
\begin{tabular}{l r r r r}
\toprule
Galaxy    & H$\alpha$ Luminosity & H$\beta$ Luminosity & \taub\ & \halphasurf\\\
AGC\#  & log(erg s$^{-1}$) & log(erg s$^{-1}$) & & log(erg s$^{-1}$ pc$^{-2}$)\\
\midrule
191702 & 38.35 $\pm$ 0.25 & 37.87 $\pm$ 0.25 & 0.07 $\pm$ 0.01 & 33.42 $\pm$ 0.31\\
212838 & 38.28 $\pm$ 0.21 & 37.78 $\pm$ 0.21 & 0.12 $\pm$ 0.02 & 33.41 $\pm$ 0.26\\
220755 & 38.11 $\pm$ 0.13 & 37.48 $\pm$ 0.13 & 0.38 $\pm$ 0.07 & 32.95 $\pm$ 0.17\\
220837 & 38.29 $\pm$ 0.13 & 37.60 $\pm$ 0.13 & 0.55 $\pm$ 0.05 & 33.34 $\pm$ 0.17\\
220860 & 38.83 $\pm$ 0.13 & 38.29 $\pm$ 0.13 & 0.20 $\pm$ 0.01 & 33.83 $\pm$ 0.17\\
221000 & 39.20 $\pm$ 0.13 & 38.61 $\pm$ 0.13 & 0.30 $\pm$ 0.01 & 33.80 $\pm$ 0.16\\
225852 & 38.34 $\pm$ 0.13 & 37.86 $\pm$ 0.13 & 0.06 $\pm$ 0.03 & 33.15 $\pm$ 0.16\\
227897 & 38.14 $\pm$ 0.13 & 37.67 $\pm$ 0.13 & 0.02 $\pm$ 0.04 & 33.11 $\pm$ 0.16\\
221004 & 38.62 $\pm$ 0.13 & 38.10 $\pm$ 0.13 & 0.15 $\pm$ 0.03 & 32.98 $\pm$ 0.16\\
202218 & 38.98 $\pm$ 0.11 & 38.37 $\pm$ 0.11 & 0.34 $\pm$ 0.03 & 33.63 $\pm$ 0.14\\
225882 & 38.97 $\pm$ 0.06 & 38.54 $\pm$ 0.06 & -0.07 $\pm$ 0.01 & 33.13 $\pm$ 0.11\\
\bottomrule
\end{tabular}
\begin{tablenotes}[para,flushleft]
Integrated values obtained from the summed spectrum for each galaxy.
\end{tablenotes}
\end{threeparttable}
\end{table*}

We plot the spatial mapping of the H$\alpha$ luminosity in Figure \ref{h_alpha}.  We find integrated H$\alpha$ luminosity values (Table \ref{results_table}) ranging from 38.01 to 39.19 log(erg s$^{-1}$), and individual spaxel H$\alpha$ luminosities ranging from 35.27 to 38.02 log(erg s$^{-1}$).  We find that H$\alpha$ emission is not uniform throughout the ionization regions, but tends to be concentrated within certain sections of the regions.  This is consistent with the picture of clumpy star formation within these dwarf irregular galaxies.

If we convert the H$\alpha$ flux using the \citet{Hao:11} calibration between H$\alpha$ luminosity and star formation rate, we would find, without correcting for dust, that the integrated galaxy SFRs are in the range -3.26 \lt log(M$_\odot$yr$^{-1}$) \lt -2.08.  With dust corrections applied, the SFRs range from -2.88 \lt log(M$_\odot$yr$^{-1}$) \lt -1.65 (Table \ref{paper2_table}).  The uncorrected SFRs are near the luminosity limit for this SFR estimation method.  \citet{Kennicutt:12} cautions that below 38 log(erg s$^{-1}$) the accuracy of the estimation will degrade as small-number statistics begin to dominate.  Therefore, we do not convert our individual spaxel H$\alpha$ luminosities into SFRs, but it is worth noting that our lowest luminosity spaxel, 35.27 log(erg s$^{-1}$), is roughly equivalent to the ionizing flux of a single B0V star \citep{Smith:02}.

\begin{figure*}
\epsfig{ file=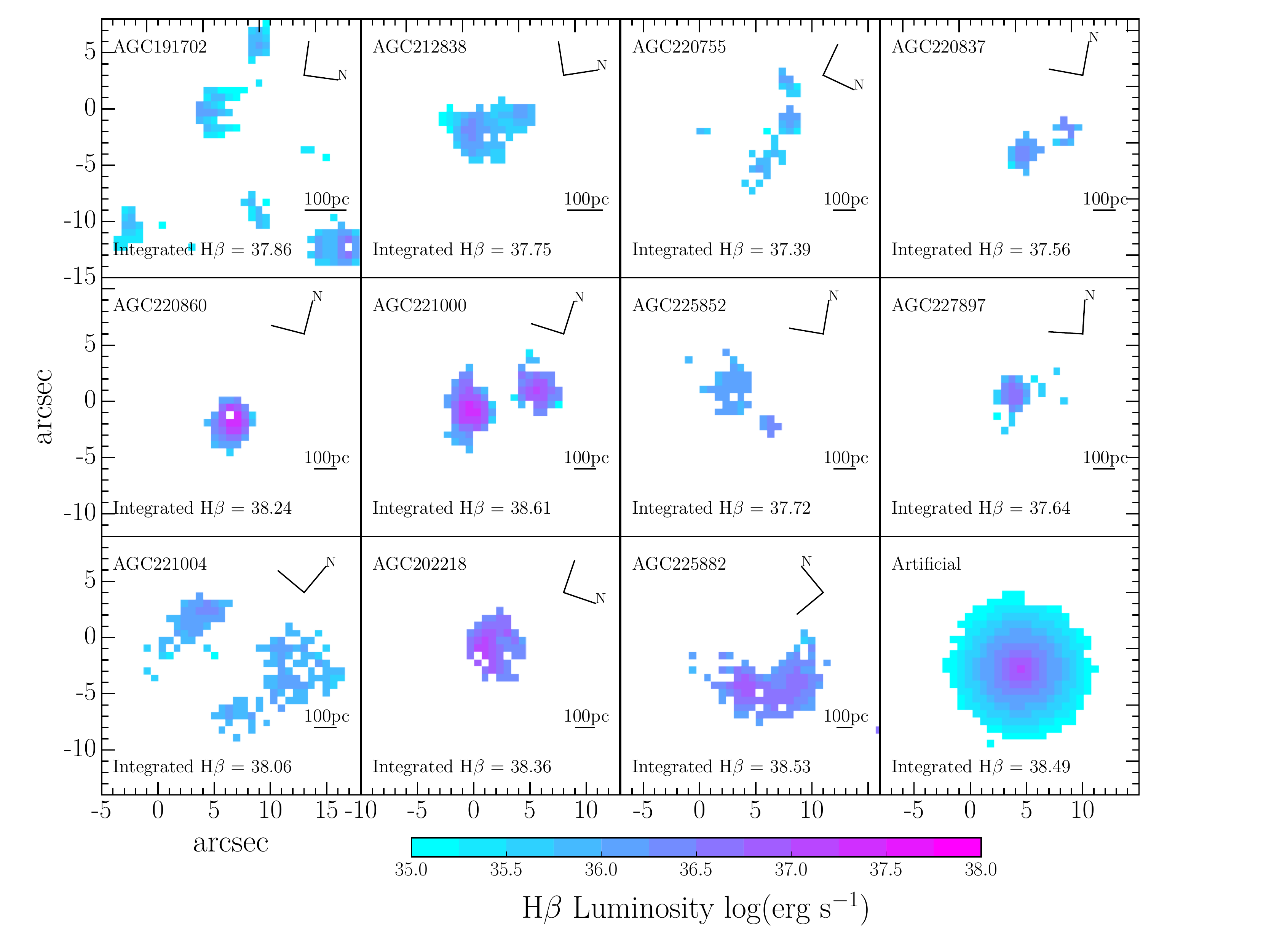, scale=0.6}
\caption[Spatial Map of H$\beta$ Luminosity]{Spatial mapping of the H$\beta$ luminosity for each galaxy.  Inset within each frame is the log H$\beta$ luminosity (erg s$^{-1}$) measured when summing the spaxels together to form a single effective spectrum.  Peaks in the H$\beta$ emission appear to be preferentially located near the centroids of the emission line regions.  Also shown in the lower right panel is the artificial galaxy created with an exponential decline in H$\beta$ luminosity as a function of radius.}
\label{h_beta}
\end{figure*}

We also show the H$\beta$ luminosity within each spaxel, calculated from the H$\beta$ line flux and luminosity distance.  
Once again, no dust corrections have been applied to the luminosity values reported.  We observe that H$\beta$ luminosity tends to peak near the center of each individual emission line region.  The peak of H$\beta$ flux values are spatially well correlated with the peaks in the H$\alpha$ spatial maps as would be expected.  We find integrated H$\beta$ luminosity values (Table \ref{results_table}) ranging from 37.39 to 38.61 log(erg s$^{-1}$), and individual spaxel H$\beta$ luminosities ranging from 34.31 to 37.41 log(erg s$^{-1}$).  We see in Figures \ref{h_alpha} and \ref{h_beta} that the artificial galaxy exhibits the radial decline in flux that would be expected based on its construction.

\subsection{Balmer Optical Depth}

\begin{figure*}
\epsfig{ file=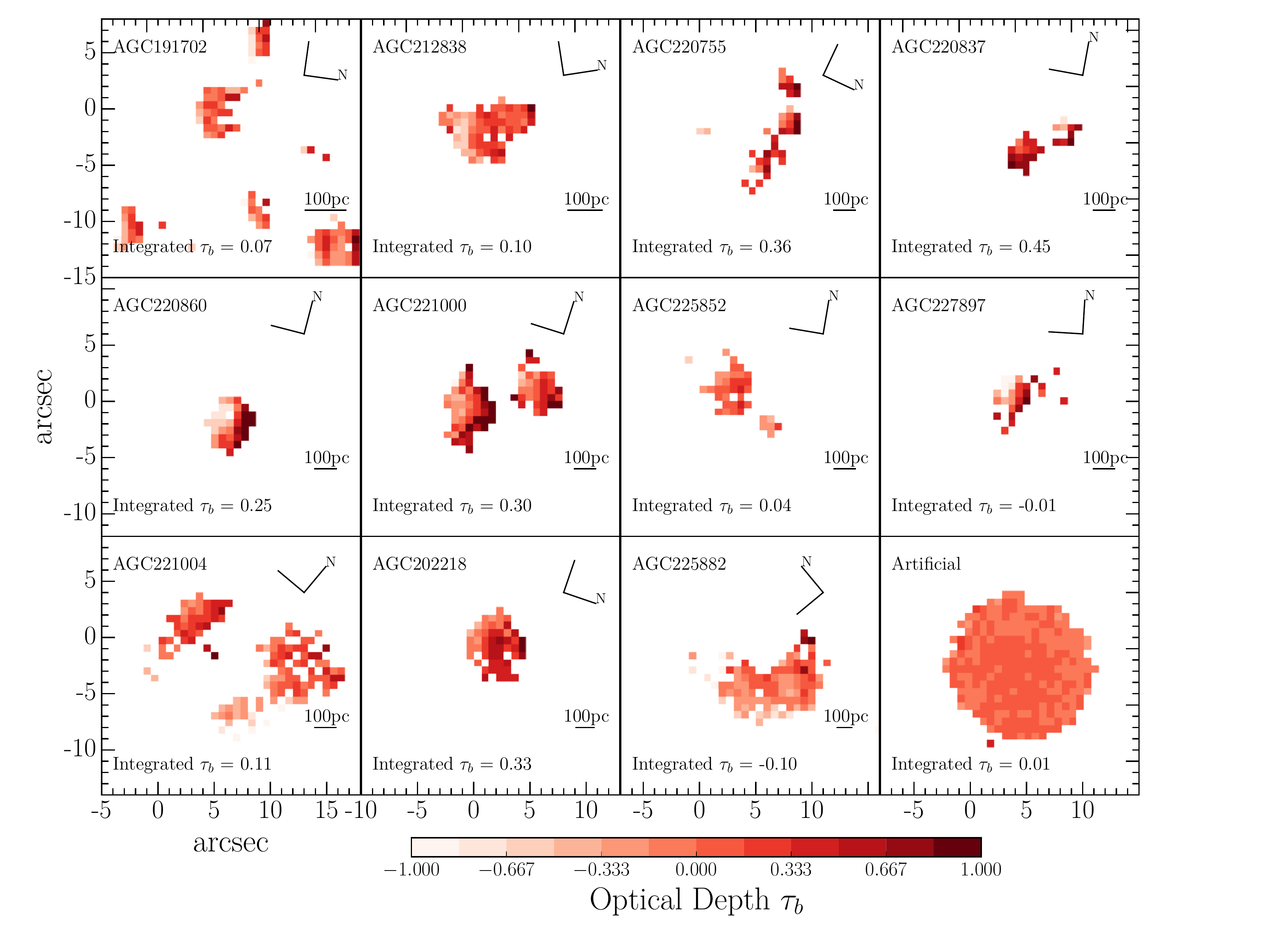, scale=0.6}
\caption[Spatial Map of Balmer Optical Depth]{Spatial mapping of the Balmer optical depth for each galaxy.  Inset within each frame is the Balmer optical depth measured when summing the spaxels together to form a single effective spectrum.  Also shown in the lower right panel is the artificial galaxy for which every spaxel was created to have a Balmer optical depth \taub\ = 0.01. The observed galaxies show greater variation within their individual spaxels than the artificial galaxy, indicating that the observed scatter is not due to noisy spaxels.}
\label{balmer}
\end{figure*}

As in \citet{Calzetti:94} and \citet{Reddy:15}, we utilize the Balmer optical depth (\taub ) defined in Equation \ref{taub}, which is the difference in optical depths measured at the H$\alpha$ and H$\beta$ wavelengths.  Therefore, \taub\ is a measurement of the degree of reddening within a region, and is related to the color-excess E(B-V).  { Assuming Case B recombination at T = 10$^4$ K with an electron density n$_e = 10^2$ cm$^{-3}$, one would find a theoretical minimum value of \taub\ = 0 \citep{Osterbrock:89}.}

We use these measurements of the optical depth to spatially map the dust content within each galaxy in Figure \ref{balmer}.  We also see in Figure \ref{balmer} that the recovered \taub\ values for the artificial spaxels produce a dust map that is flat, as it was constructed to be.  This is in contrast to the dust distribution in the observed galaxies, which is patchy and irregular.  We also provide an integrated \taub\ value inset within each image frame, which is obtained by summing the spectra in each spaxel that passes the AoN cut, and measuring the line ratios of the combined effective spectrum.  This is roughly equivalent to the Balmer optical depth of the entire galaxy if it were to be observed in a single fiber.  It appears that the spaxels with higher H$\alpha$ luminosity correlate with the spaxels containing higher Balmer optical depths (Figures \ref{h_alpha} \& \ref{balmer}).

We measure integrated Balmer optical depths ranging from { -0.07 \lt \taub\ \lt 0.55}.  This range is much smaller then that of observed L* galaxies at $z=2$ in \citet{Reddy:15}, where they find galaxies in the range -1.5 \lt \taub\ \lt 1.0.  On a spaxel-by-spaxel basis, we find \taub\ values ranging from -1.33 to 1.64.

%

\subsection{Correlation Between H$\alpha$ and \taub }

\begin{table}
\centering\footnotesize
\caption{Correlation Between \halphasurf\ and \taub\ in Dwarf Galaxies}
\label{fits_table}\medskip
\begin{threeparttable}
\begin{tabular}{l r r r r r}
\toprule
Galaxy & Slope & Y-Intercept & $\rho$ & Std Dev\\
 &  &  & & from \\
 &  &  & & Artificial\\
\midrule
191702* & 0.27 $\pm$ 0.07 & -9 $\pm$ 2 & 0.39 & 1.5\\
212838 & 0.28 $\pm$ 0.10 & -9 $\pm$ 3 & 0.34 & 0.8\\
220755* & 0.61 $\pm$ 0.24 & -20 $\pm$ 8 & 0.45 & 1.3\\
220837 & 0.73 $\pm$ 0.35 & -24 $\pm$ 12 & 0.24 & -0.1\\
220860* & 0.33 $\pm$ 0.16 & -11 $\pm$ 5 & 0.42 & 1.0\\
221000 & 0.12 $\pm$ 0.07 & -4 $\pm$ 2 & 0.17 & -0.8\\
225852* & 0.70 $\pm$ 0.18 & -23 $\pm$ 6 & 0.42 & 1.1\\
227897 & 0.18 $\pm$ 0.23 & -6 $\pm$ 8 & 0.16 & -0.5\\
221004* & 0.82 $\pm$ 0.10 & -27 $\pm$ 3 & 0.65 & 5.8\\
202218 & 0.52 $\pm$ 0.14 & -17 $\pm$ 5 & 0.36 & 0.8\\
225882* & 0.37 $\pm$ 0.10 & -12 $\pm$ 3 & 0.34 & 1.0\\
Artificial & 0.03 $\pm$ 0.01 & -1 $\pm$ 1 & 0.25 & 0.0\\
\\
All Spaxels* & 0.26 $\pm$ 0.03 & -8 $\pm$ 1 & 0.39 & 4.0\\
Integrated & 0.23 $\pm$ 0.18 & -7 $\pm$ 6 & 0.20 & -0.2\\
\bottomrule
\end{tabular}
\begin{tablenotes}[para,flushleft]
Galaxies with an asterisk by their name indicate those for which the correlation between \halphasurf\ and \taub\ is significant.
\end{tablenotes}
\end{threeparttable}
\end{table}

\begin{figure*}
\epsfig{ file=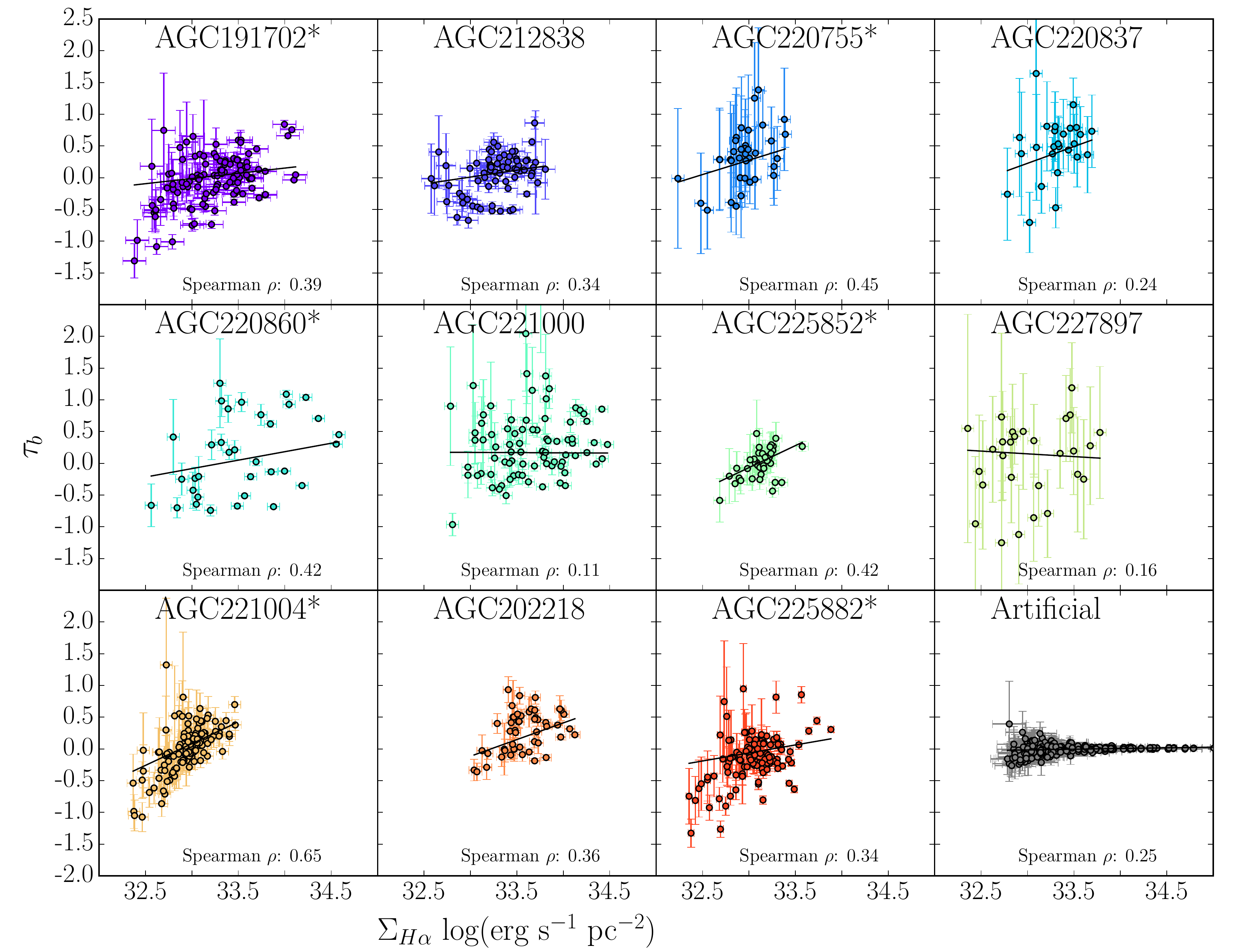, scale=0.55}
\caption[\halphasurf\ vs. \taub\ for Individual Galaxies]{H$\alpha$ surface density and Balmer optical depth for individual spaxels within each galaxy.  There exists a { positive correlation ($\geq$ 1.0$\sigma$ significance) between H$\alpha$ luminosity and \taub\ for 6 out of 11 galaxies.  Galaxies which have passed this threshold for significance are indicated by an asterisk following the galaxy name.}  The artificial galaxy, which was constructed to have a flat \taub\ profile, exhibits no correlation.}
\label{h_alpha_vs_balmer}
\end{figure*}

\begin{figure}
\epsfig{ file=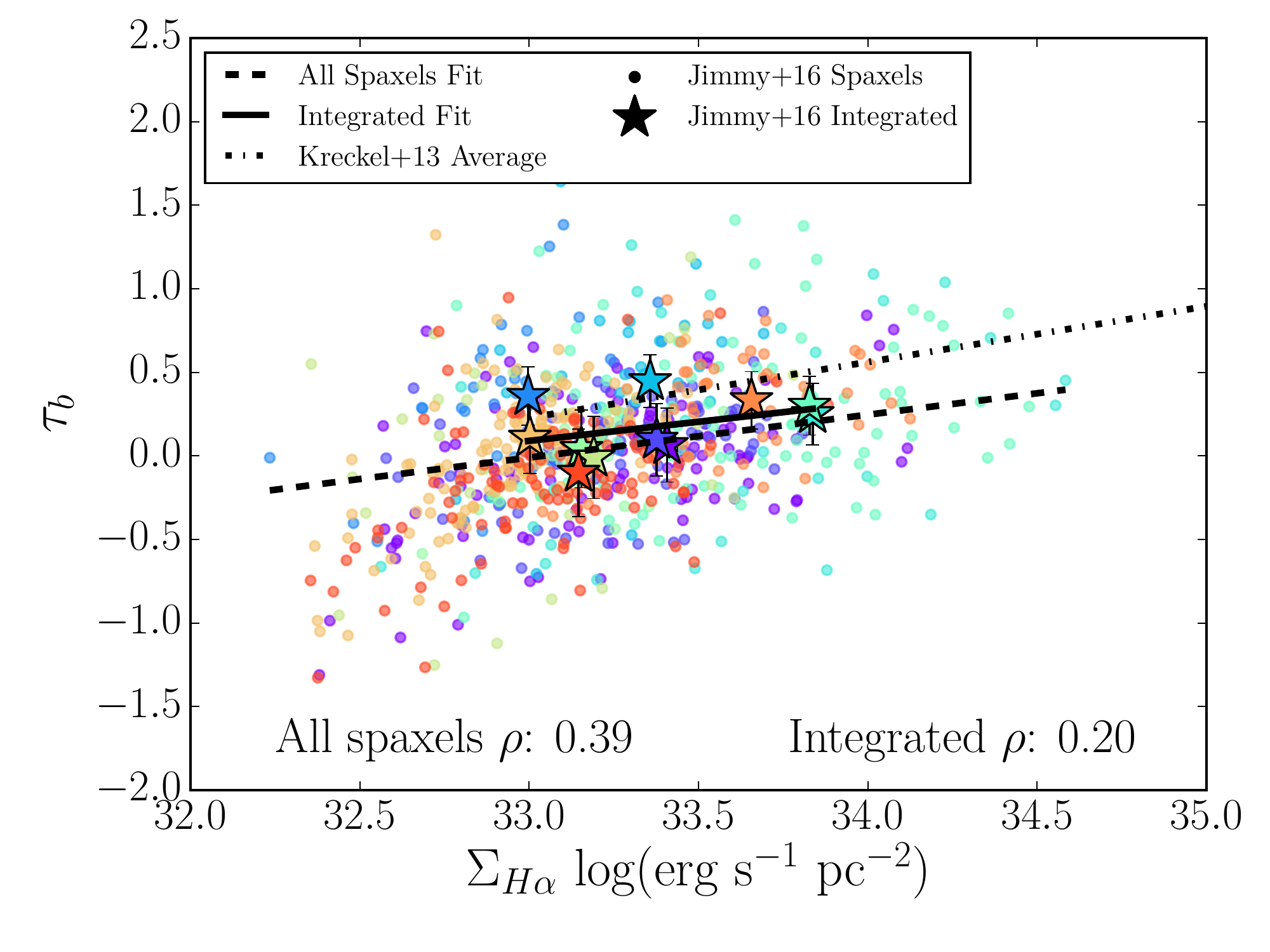, scale=0.45}
\caption[\halphasurf\ vs. \taub\ All Spaxels Fit]{H$\alpha$ surface density and Balmer optical depth for the individual spaxels (dots) using the same color scheme as Figure \ref{h_alpha_vs_balmer}.  Also plotted is the mean of the relations found in \citet[dot-dash]{Kreckel:13} which has a larger spatial scale, and shows a very similar slope.  For comparison, we have included the integrated values of each galaxy (stars).  The slope of the integrated values fit is similar to that of the fit to the individual spaxels.}
\label{h_alpha_vs_balmer_all_in_one}
\end{figure}

We test to see if H$\alpha$ luminosity and the Balmer optical depth are correlated on a per-spaxel basis, as well as (using the integrated fluxes) a galaxy-by-galaxy basis.  To account for the effects of observing galaxies at varying distances with varying spatial resolution scales, we divide each measurement by the surface area to show the H$\alpha$ luminosity surface density (\halphasurf) in Figure \ref{h_alpha_vs_balmer}.  The H$\alpha$ surface density is defined as \halphasurf = $L_{\text{H}\alpha}$/(s$^2 \cdot $n) where $L_{\text{H}\alpha}$ is the H$\alpha$ luminosity, s is the physical size of the spaxel in pc, and n is the number of spaxels in the bin.  We find that there is a general trend towards spaxels with higher \halphasurf\ having higher \taub\ values.  We fit the spaxels using the curve\_fit routine from SciPy to perform a weighted least-squares linear fit to the individual spaxels in Figure \ref{h_alpha_vs_balmer} and report our results in Table \ref{fits_table}.  The curve\_fit routine determines the best fit by minimizing the sum $\Sigma ((f(x)-y)/w)^2 $ where f(x) is the function being fit, y is the y-axis values, and w is the uncertainties in the y-axis values.  Using a linear function as f(x), all galaxies exhibit either a flat or positive slope.  In Figures \ref{h_alpha_vs_balmer} and Table \ref{fits_table}, the slope of the artificial galaxy differs significantly from the slopes of the observed galaxies, indicating that low S/N spaxels are not primarily driving the apparent relation between \halphasurf\ and \taub .

We test for correlation using Spearman's rank correlation coefficient ($\rho$), as provided by the SciPy routine spearmanr.  Similar to the Pearson correlation coefficient, the Spearman coefficient can range in value from -1 to 1.  We use the Spearman coefficient because it does not require that the distribution of the data be Gaussian in nature, like the Pearson correlation coefficient does.  Our X-axis values are not Gaussian in that there are generally more low luminosity spaxels than high luminosity spaxels.

Typically, the following criteria are used to determine the significance of $\rho$ values: $0.10 <  \rho < 0.29$ is a weak association, $0.30 <  \rho < 0.49$ is a moderate association, $\rho > 0.50$ is a strong association \citep{cohen:88}.  The artificial galaxy was constructed to have no correlation between the two values, and exhibits a $\rho$ = 0.25, consistent with a weak correlation, and indicative that these significance thresholds are accurate.  Based on the low Spearman rank coefficient in the Artificial galaxy, we conclude that using H$\alpha$ to calculate \taub\ is not the primary driver of the relation between H$\alpha$ and \taub .  

In addition, we use the $\rho$ value found for the Artificial galaxy to perform a null-hypothesis test, where we use the Spearman coefficient of the Artificial galaxy as our null { hypothesis}, and determine the number of standard deviations each galaxy's $\rho$ measurement is from the Artificial galaxy.  Performing a Fisher z-transformation on both the { artificial} data and the observed data allows us to determine the significance of the separation as { $\sigma = $(z($\rho$)-z($\rho_0$))*$\sqrt{n-3}$}.  We report these standard deviations from the null hypothesis as the final column in Table \ref{fits_table}.  The factor of $\sqrt{n-3}$ in the numerator indicates that the statistical significance of the deviation from the control is dependent upon the number of spaxels (n) for a given galaxy.  

{ 8 of 11 galaxies show a positive deviation from the null hypothesis, however to ensure our results are robust we utilize a $\geq$ 1.0$\sigma$ threshold for significance.}  Based on this criteria, we find for { 6 out of 11} galaxies that the spaxels have a statistically significant correlation between the observed H$\alpha$ luminosity surface density and the Balmer optical depth.  These galaxies are indicated with an asterisk in Table \ref{fits_table}.  There does not appear to be any correlation between galaxy distance and $\rho$, indicating that our results are not dependent upon spatial resolution.  { The 5 galaxies that do not appear to follow the positive correlation between \halphasurf\ and \taub\ (AGC202218, AGC212838, AGC220837, AGC221000, and AGC227897) do not appear to have any particular commonalities}.  They vary widely in stellar mass, { HI-gas mass,} and metallicity as can be seen in Table \ref{paper2_table}.  

The number of spaxels for each galaxy ranges from 26 to 130, so in some galaxies such as AGC220837 and AGC227897 there may not be enough spaxels to reliably derive a slope to this relation, as is evident by the low significance of the deviation from the Artificial galaxy.  Therefore, we combine all spaxels from all of the galaxies to form Figure \ref{h_alpha_vs_balmer_all_in_one}.  When performing a linear fit we find a statistically significant correlation between the two values with a slope of 0.26$\pm$0.03 (Table \ref{fits_table}).  The Spearman rank coefficient of the aggregate spaxels is 0.39, higher than the Artificial galaxy, { which would be catagorized as a moderate correlation.  The Spearman rank coefficient of the aggregate spaxels is 4$\sigma$ offset from the Artificial galaxy, indicating a high statistical significance of our result.}

We also provide the results of a linear fit to the integrated values.  The slope found for the linear fit to the integrated values for each galaxy is similar to that of the individual spaxels.  With only 11 points in the integrated values fit, it is difficult to draw any robust conclusions, as is indicated by the low significance (0.2$\sigma$) for the integrated spaxels.  Although we must caution that the low sigma value does not indicate that the relation itself is insignificant, rather that the number of data points we're using to calculate the Spearman coefficient is low, and therefore the factor of $\sqrt{n-3}$ will be low. 

\vspace{10 mm}

\section{Discussion}

\subsection{Increasing Attenuation with Increasing H$\alpha$ Luminosity}

The correlation between enhanced reddening and higher H$\alpha$ luminosity within individual spaxels indicates that dustier regions correlate with higher SFR regions.  This correlation has been observed in larger galaxies \citep{Kreckel:13}, but we have shown for the first time that it holds true for gas-rich dwarf galaxies (Figure \ref{h_alpha_vs_balmer}).  Our slope is within 2$\sigma$ of the mean of the \citet{Kreckel:13} data (0.32, Figure \ref{h_alpha_vs_balmer_all_in_one}), which was focused on emission line regions within more massive spiral galaxies.

\citet{Kreckel:13} predicted, based on the lack of high $\Sigma_{SFR}$ regions having low A$_V$, that HII regions are likely to be correlated with the dust on 20-120 pc scales, but the physical resolution of their data was limited to 100 pc.  Using our 30-80 pc resolution IFU spectroscopy, we are able to confirm their hypothesis that emission line regions and dust are well correlated on those scales for { 6 of 11 galaxies (when applying our more conservative threshold of $\geq$ 1$\sigma$ significance)}.  It is unclear at which physical scale this relation will break down.  \citet{Kreckel:13} also compared emission line regions to stellar continuum as dust tracers and concluded that emission line regions were a better tracer.  We are able to investigate effects smaller than 100pc in diameter, the nominal size of the giant molecular clouds that form small clusters of stars.  However, we are unable to probe down to sizes of $\sim$10 pc, the size of cores that form individual stars \citep{Tan:15}. 

It is likely that the correlation between \halphasurf\ and \taub\ is caused by the young stars within these emission line regions being surrounded by the dusty birth-cloud in which they formed.  If dust obscuration traces instantaneous star formation, then the physical effect causing this must be acting on similar time-scales.  The cloud of gas and dust that the young star formed from will not be entirely consumed in the formation of the star, and it takes time (approximately 30 Myrs, on par with the lifetime of an O or B type star) for the dust cloud to dissipate \citep{Charlot:00}.


We observe that the slope is similar, within the uncertainties, between the spaxel-by-spaxel relation and the integrated values relation.  The Spearman $\rho$ for the integrated galaxies is not significantly different from the Artificial galaxy we used to formulate our null hypothesis.  Other studies (e.g. \citealt{Reddy:15}) have shown that dustier galaxies have higher H$\alpha$ luminosity on integrated galaxy-wide scales.

Our results agree with the conclusions of \citet{Kreckel:13} that emission line sources are preferentially located within dusty birth clouds.  Similarly \citet{Reddy:15} found that their results were consistent with a two population model.  Firstly a modestly reddened underlying stellar population, with a second dustier stellar population that begins to dominate the nebular line luminosity with increasing SFR.  Our results align with this view, although it would be advantageous to compare our emission line measurements to stellar continuum measurements, we do not have the UV observations that would be necessary to properly constrain our data.  Based on the spatial agreement between the peaks in the Balmer optical depths and the H$\alpha$ luminosity, it is likely that the star formation rates within our dwarf irregular galaxy population are not powerful enough to have a well-mixed ISM for the majority of our sample.   We note that galaxies with super strong outflows, like NGC 2146 from \citet{Kreckel:13} exhibit spatial offsets in their dustiest regions and their highest H$\alpha$ regions.  NGC 2146 also shows near agreement between stellar continuum reddening and gas emission line reddening, suggesting that stellar winds have blown away the clouds of gas and dust and the emission lines experience a grayer dust attenuation.

\subsection{Negative \taub\ Values}
As can be seen in Figures \ref{balmer}, \ref{h_alpha_vs_balmer} \& \ref{h_alpha_vs_balmer_all_in_one}, a large number of the spaxels within each galaxy exhibit \taub\ values that are seemingly unphysical.  It would appear that either H$\alpha$ is being over-estimated or H$\beta$ is being underestimated.  The following possibilities were considered as the cause of the negative \taub\ values observed:
\begin{itemize}

\item Poor Gaussian Fits: We have checked all of the fits to each spaxel spectrum to verify that the Gaussians do indeed provide a reasonable fit to the data, and find no obvious source of error from mis-measurement of the flux values.  We also checked to ensure that our de-convolution of the blended H$\alpha$ and [NII] emission lines was not a cause by fixing the ratio of the three lines to the integrated value and re-measuring all of the spaxels with no significant difference in our results.  Even if that were a factor, it would only serve to increase the H$\alpha$ flux with any corrections, and that would further exacerbate the issue.

\item Shocks of the ionized gas would cause a change in the [NII]/H$\alpha$ ratio, however our Python-based Gaussian fitting routines allow the ratio between H$\alpha$ and [NII] to vary on a spaxel-by-spaxel basis.

\item { Line Sensitivity: We have also observed in our artificial galaxy that low flux spaxels are not significantly lower in \taub\ (Figure \ref{h_alpha_vs_balmer}).  The artificial galaxy was given an input \taub\ = 0.01 for each spaxel, and of the 111 total spaxels recovered for the artificial galaxy, 53 have \taub\ $ > 0.01$ and 58 exhibit \taub\ $ < 0.01$ indicating no significant bias towards lower \taub\ values.  We also find no significant correlation between AoN of each spaxel and \taub , therefore it is unlikely that the negative \taub\ values observed are due to a line-sensitivity effect.}

\item Instrumental Effects: We created a field of random emission lines within a blank data cube to confirm that there are not issues with certain spaxels, or gradients across the field-of-view or CCD. 

\item Balmer Absorption: We also considered that our assumption of Balmer absorption being negligible was at fault, however, Balmer absorption more strongly affects H$\alpha$ than H$\beta$, so any corrections we apply to account for absorption will create only lower \taub\ values.

\end{itemize}

We are therefore reasonably confident that the \taub\ values we are measuring are accurate representations of the observations.

Typically studies will assume that a ratio of H$\alpha$/H$\beta$ < 2.86 is indicative of little to no dust within a galaxy, and for the purpose of their analysis treat negative Balmer optical depths as 0 (e.g. \citealt{Reddy:15}).  Another typical approach is to state that the intrinsic ratio between H$\alpha$ and H$\beta$ varies as a function of electron temperature and density (e.g \citealt{Battisti:16}).  However, as we will show, this is typically a small effect, and insufficient to explain all negative \taub\ values.  Due to the large number of individual spaxels within our sample with \taub\ < 0 {( 43\%)}, we cannot ignore, or make the ``no dust'' assumption, for all of negative \taub\ spaxels.  This then leads us to speculate on the cause of the negative \taub\ values.

The minimum value of 2.86 for the ratio of H$\alpha$ to H$\beta$ from \citet{Osterbrock:89} assumes three important factors, Case B recombination (meaning that the gas is optically thick to Lyman alpha photons), electron temperature T$_e$ = 10,000K, and electron density n$_e = 10^2$ cm$^{-3}$.  Changing our assumption to Case A recombination, in which Lyman alpha photons can readily escape, is likely not a solution as such a galaxy would be very low surface brightness and difficult to observe \citep{Aller:84}.  If we continue to assume Case B recombination and allow the temperature to vary, we find that increasing the assumed temperature can lower the ratio, such that if we double the temperature to T$_e$ = 20,000K, we find H$\alpha$/H$\beta$ = 2.76.  If we were to additionally alter our assumption and increase the density of electrons to n$_e = 10^6$ cm$^{-3}$ we would find again only a marginal effect, with the new theoretical minimum H$\alpha$/H$\beta$ = 2.72.  

{ Under these new assumptions, 30\% of our sample would continue to exhibit a negative \taub\ value.  Therefore assuming a different temperature and density is not sufficient to explain all of the negative \taub\ values observed.  It is however still possible that the assumption of Case B recombination may not be appropriate for all galaxy studies.}

{ As the Balmer decrement is only moderately sensitive to the physical assumptions being made}, we speculate that the possible cause of { a portion of the} negative \taub\ values may be due to scattering of blue light from within the dusty birth clouds to regions of the galaxy that have less dust, and also fewer young O and B type stars capable of producing the ionizing radiation to induce the Balmer series of emission lines.  Regions that have less Balmer series emission overall would be more sensitive to even small increases of scattered H$\beta$ flux originating from within the very active and dusty regions.  { As this flux would then be allowed to escape, there would not be an equal contribution of H$\beta$ flux returning to the dustier regions.}

{ In future observations, longer integration times and higher spectral resolution observations would provide stellar continuum data.  Modeling the stellar continuum would allow us to properly address the Balmer absorption.  Additionally, higher resolution observations would provide more robust deblending of the H$\alpha$ and [NII] emission lines.  These improvements in observations would provide greater confidence in our measurement of negative \taub\ values.}

\vspace{10 mm}

\section{Conclusion}

We have continued the study of 11 IFU observed dwarf galaxies first reported on in \citet{Jimmy:15}, where they studied the fundamental metallicity relation as a function of HI-gas mass and star formation rate.  This sample was selected to have low HI-gas mass (M$_{HI}$ \lt 10$^{8.15}$ M$_\odot$), and as such exhibits low stellar-mass, low metallicity, and low SFR (Table \ref{paper2_table}).  We continued our observations within this work by studying the H$\alpha$ and H$\beta$ luminosity, in addition to the Balmer optical depth (\taub ).  We measured integrated H$\alpha$ luminosities ranging from 38.01 \lt log(H$\alpha$) \lt 39.19 erg s$^{-1}$ and integrated \taub\ values in the range { -0.07 \lt \taub\ \lt 0.55} (Table \ref{results_table}).

Dust curves are used to correct for the attenuation of light when determining star formation rates.  This attenuation is affected by the distribution of dust within the ISM of galaxies, as well as the composition and size of the dust grains themselves.  Gas-rich star forming galaxies are likely to have their bluest star-forming regions surrounded by dusty birth clouds, which will preferentially scatter and absorb the blue light, causing a steeper dust attenuation law of the ionized gas emission, as observed in the Small Magellanic Cloud.

Utilizing the spatial information provided by IFU spectroscopy, we mapped the reddening properties of these dwarf irregular galaxies (Figure \ref{balmer}).  We found that the regions of the galaxy with more reddening positively correlate with regions of enhanced H$\alpha$ emission.  This correlation has been demonstrated before in larger spiral galaxies \citep{Kreckel:13}, however, we have shown that the correlation between SFR and dust in emission line regions likely continues to hold even in gas-rich dwarf galaxies.  Using Spearman's coefficient, we quantified the degree of correlation between \halphasurf\ and \taub .  We found { 8 of 11 galaxies showed a statistically significant positive correlation at $\geq$ 0.8$\sigma$ (6 of 11 at $\geq$ 1.0$\sigma$) between these two values}.  If we test for a correlation within all spaxels from all galaxies at once, we find $\rho = 0.39$ with 4$\sigma$ significance in deviation from our null hypothesis derived from the Artificial galaxy (Table \ref{fits_table}).  We demonstrated that this correlation continues to hold in low mass galaxies, down to physical scales as low as 30pc.  Higher spatial resolution observations of similar dwarf galaxies would help determine the physical resolution at which this relationship breaks down.

We also discussed the possibility that the negative \taub\ values we measured may be indicative of low-luminosity regions having blue light scattered into them, causing seemingly unphysical \taub\ measurements within some of our spaxels.  { Alternatively, it may be that assuming Case B recombination for this population of galaxies is inappropriate.}

This is the first study to specifically target local universe, gas rich, dwarf irregular galaxies for a correlation between \halphasurf\ and \taub .  Because these galaxies are local analogs to the high-redshift galaxies currently being observed, we have shown that assuming a simple uniform dust distribution within high-redshift star forming galaxies is likely to be inaccurate.  
The clumpiness of our { Balmer optical depth maps (Figure \ref{balmer})} would suggest an SMC type reddening curve is most appropriate for low-metallicity dwarf irregular galaxies.  Additionally our dwarf irregular galaxies have similar SFRs, stellar-masses, and metallicity as the SMC.  Although it is important to note that we are unable to resolve individual stars with our observations, so it is possible that other physical effects such as grain size or composition that we are unable to explore may reproduce these same conclusions.  

Assuming that young star forming galaxies at high redshift also have their star forming regions surrounded by dusty birth-clouds, we would argue for a steeper SMC-like dust law for those galaxies as well.


\acknowledgments
\section*{Acknowledgments}

{ We would like to thank the referee for the careful review and the valuable comments, which provided insights that helped improve the paper.}  We are very grateful to the entire ALFALFA team for the extensive work involved in observing, processing, flagging, and cataloguing the ALFALFA data, which this paper is based on. We also wish to acknowledge the important contributions of Martha Haynes, Riccardo Giovanelli, and Marco Scodeggio in devising the VIMOS-IFU follow-up project based on ALFALFA-detected dwarf galaxies.  SB acknowledges the funding support from the Australia Research Council through a Future Fellowship (FT140101166).  POA acknowledges ARC Discovery Project DP130101460.

\bibliographystyle{apj}
\bibliography{Jimmy2016}

\end{document}